\documentclass[%
 reprint,
 amsmath,amssymb,
 aps,
floatfix,
]{revtex4-2}

\usepackage{graphicx}% Include figure files
\usepackage{dcolumn}% Align table columns on decimal point
\usepackage{bm}% bold math
\usepackage{adjustbox}
\usepackage{tikz}
\usetikzlibrary{shapes.geometric, arrows}
\usepackage{url}
\usepackage{xcolor}
\usepackage[normalem]{ulem}

\begin{document}

\preprint{APS/123-QED}

%\title{Nonlinear dynamics optimization with multi-objective multi-generation Gaussian process optimizer for SPEAR3 upgraded lattice}% Force line breaks with \\
\title{Correction of storage ring optics with an improved closed-orbit modulation method}
% Force line breaks with \\
\author{Xiaobiao Huang}
\email[]{xiahuang@slac.stanford.edu} 
\affiliation{SLAC National Accelerator Laboratory, Menlo Park, CA 94025, USA}
\author{Xi Yang}
\email[]{xiyang@bnl.gov} 
\affiliation{Brookhaven National Laboraotry, Upton, USA}
%\affiliation{%


%\affiliation{
%\affiliation{
%\affiliation{%


\date{\today}% It is always \today, today,

\begin{abstract}
We improved a previously proposed method of using closed-orbit modulation for linear optics correction. 
Instead of fitting individual closed orbits, the improved method decomposes the orbit oscillation data into 
two orthogonal modes and fits the amplitudes of the modes at all BPMs. 
While the original method is limited to process around tens to a hundred orbits, the improved method can process 
thousands of orbits, which are easily available when alternating-current (AC) waveforms are applied to the two modulating correctors.
The method has been experimentally demonstrated on the National Synchrotron Light Source (NSLS)-II storage ring. 
\end{abstract}

\maketitle

%\tableofcontents

\section{Introduction}
Beam-based correction for storage ring linear optics  is critical for storage ring commissioning and operation. 
This is a well researched area. 
Since the first successful demonstration of global optics correction with the linear optics from closed orbit (LOCO) method~\cite{SAFRANEK199727}, there have been many other methods in the literature. 

Almost all methods rely on utilizing the optics information hidden in the closed orbit or turn-by-turn (TBT) orbit 
measurements. 
The TBT orbit based methods~\cite{HuangICA05,HuangFitTbTPRSTAB10,AibaLHC1stOptics,TomasLHCOptics2010,ShenRHIC2013,YangICANSLS2}
have the advantage of fast data acquisition and sampling the full betatron phase space, 
but also have the disadvantages of requiring kicker or pinger magnets in both 
transverse planes and suffering from decoherence of the kicked motion. 
AC-LOCO, the approach of taking orbit response matrix data by driving multiple correctors with alternating-curent (AC) waveforms of different 
frequencies~\cite{DiamondRFLOCO,ALBARFLOCO,XiYangACLOCO}, substantially speeds up data acquisition for the LOCO method. However, 
It requires fast correctors distributed around the ring, while in reality, often times only a fraction of all correctors are fast
and they are typically located at the few identical locations in each cell, limiting the sampling points. 

The linear optics from closed orbit modulation (LOCOM) method~\cite{LOCOM}, proposed recently, enjoys the advantages of 
the previous methods 
in both the closed-orbit and TBT orbit camps and is spared of their disadvantages. 
The method uses two correctors to modulate the closed orbit and fit the modulated orbit to the lattice in the same fashion 
as LOCO for optics errors. 
Data acquisition is fast, while no special magnets are needed. A full sampling of the betatron phase space can be 
achieved by simply properly choosing the waveform phases of the two correctors. 
It does not suffer from orbit decoherence, hence suitable for cases with high chromaticity and high amplitude dependent 
detuning, which are common for newer, high performance storage rings.

In this study, we propose a major improvement to the LOCOM method in how the closed-orbit data are processed. 
As the original LOCOM method fits the modulated orbits, it can utilize only tens or slightly over a hundred orbits, as limited by 
the computing resource needed for least-square fitting, especially for large rings. 
Instead of fitting the raw orbits, the improved method extracts features of the modulated orbits and only fits these features 
to the model. 
The improved method is ideal for the cases when orbit modulation is done by driving AC waveforms on fast correctors, in such 
cases thousands of orbits can be taken within a second. 
This method has been demonstrated in experiments at the  NSLS-II storage ring, where the optics correction results can be 
independently verified with TBT orbit measurements. 

In the next sections, we will first describe the improved LOCOM method (Section~\ref{secMethod}), which is followed by a description of the 
 experimental application on NSLS-II (Section~\ref{secAppNSLS2}. 
The conclusion is given toward the end of the paper. 

\section{The improved LOCOM method \label{secMethod}}
For the LOCOM method, two correctors are used to drive the closed orbit deviations in each transverse plane in 
order to  sample the linear optics. The phase advance between the two correctors is preferred to be close to 
$\frac{\pi}{2}$ (modulo $\pi$). When the correctors are driven by sinusoidal waveforms and the two waveforms have a proper phase 
differences, the modulated closed orbit sweeps through betatron phase space approximately along the design 
ellipse~\cite{LOCOM}. The orbits corresponding to various points on the waveforms are selected for optics fitting. 
For example, we can choose many points with equal distance in phase on the waveforms. As these points are distributed 
evenly along the ellipse, they sample the linear optics effectively; the same effectiveness is achieved in LOCO through the use 
of many correctors distributed in different locations. 

Unlike LOCO, where the number of orbits that are used to sample the linear optics is limited by the number of correctors, 
LOCOM can produce as many orbits as desired. 
However, not all of orbits can be used in fitting for optics errors, as the least-square fitting method~\cite{SAFRANEK199727,HuangConstrainedLOCO} requires the calculation of the Jacobian matrix, which scales linearly with the number of orbits and can use up the memory of the computer. 
This is particularly true for larger rings with many BPMs and many quadrupole magnets. 
A large Jacobian matrix also slows down the calculation of matrix inversion needed in the fitting algorithm. 
The situation is similar to LOCO, for which 
usually only tens or just above 100 orbits are included for fitting. 
As an example, take a large ring with $N_B=560$ BPMs and $N_Q=400$ fitting quadrupole parameters, if 120 orbits are included for fitting, 
the Jacobian matrix would have about $2\times10^8$ elements, not including the cross-plane orbits and skew quadrupole 
fitting parameters. %560*2*120*(400+560*2)

\begin{figure}[thb]
\includegraphics[width=0.45\textwidth]{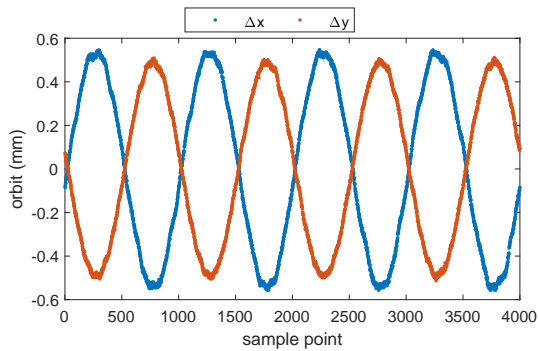}
\caption{Example of SPEAR3 AC LOCOM orbit data on one BPM, with two correctors for each plane 
driven at 4 Hz with a 4 kHz data rate for 
correctors and BPMs.} 
\label{figACorbitEx}
\end{figure}

It may be argued that including many more than 100 orbits will have diminishing returns as the orbits close in phase space 
provide similar information; the denser the orbit points on the phase space ellipse, the more the adjacent orbits look alike. 
Nonetheless, theoretically, including more orbits is preferable as it brings in more information,  in terms of both statistical merits 
and phase space diversity. 
This is critically important for the case when the correctors are driven by AC waveforms. In such a case, 
ten thousand of closed orbits may be 
measured within one second, but with less precision than the approach of sweeping through waveforms step by step (which we 
refer to as the direct current (DC) mode). 
Figure~\ref{figACorbitEx} shows an example of AC LOCOM data from the SPEAR3 storage ring, where two correctors per plane were driven with 4 Hz waveforms at a 4 kHz data rate. 
The BPM data error sigma is estimated by singular value decomposition to be about 5~um for both planes, at least 10 times 
higher than averaged orbits. 
For the original method of fitting the closed orbits directly, if 100 orbits are included for fitting, only 2.5\% of all 
orbits are used. 

In this study, we note that as the pair of correctors are driven in sinusoidal waveforms, 
the orbit observed at any BPM is also in sinusoidal waveforms of the same frequency. 
Therefore, the orbit waveforms can be decomposed into the two orthogonal oscillation modes, 
one ``sine'' mode and the other the ``cosine'' mode. 
The amplitudes of the two oscillation modes at each BPM represent the optics information 
completely. 
Following the notation in Ref.~\cite{LOCOM}, the two correctors in one plane are denoted 
correctors 2 and 1, respectively. At the downstream end of corrector 1, the position and 
angle due to the %one-pass 
kicks $\theta_2$ and $\theta_1$ are related to the optics functions 
at the two corrector locations as well as the phase advance between them. 
The position and angle vector can be decomposed as
\begin{align}
    {\bf y}_1(n) \equiv \begin{pmatrix} y_1 \\ y'_1
    \end{pmatrix} = {\bf Q}_1\begin{pmatrix} \cos 2\pi \nu n \\ \sin 2\pi \nu  n
    \end{pmatrix},
\end{align}
where the $2\times2$ matrix ${\bf Q}_1$ contains information of both the optics (Courant-Snyder parameters and 
betatron phase advance) at the two correctors and their excitation waveforms, $\nu$ represents the excitation frequency.
The elements of the ${\bf Q}_1$ matrix are easy to derive and are omitted here. 
The closed-orbit at a BPM $P$ downstream corrector 1, but before corrector 2, can be calculated with 
\begin{align}\label{eq:AmpFormMat}
    {\bf Y}_P(n) = {\bf M}_{P1}\left({\bf I} - {\bf M}_1 \right)^{-1} {\bf Q}_1\begin{pmatrix} 
    \cos 2\pi\nu  n \\ \sin 2\pi\nu  n
    \end{pmatrix},
\end{align}
where ${\bf I}$ is the identity matrix, 
${\bf M}_1$ is the one-turn transfer matrix at corrector 1, 
and ${\bf M}_{P1}$ is the transfer matrix from corrector 1 to BPM $P$. 
At the BPM, we observe the position element of vector ${\bf Y}_P$. 
By decomposing the position waveform to the sine and cosine modes, we obtain the (1,1) and (1,2)
elements of the matrix 
${\bf M}_{P1}\left({\bf I} - {\bf M}_1 \right)^{-1} {\bf Q}_1$, which are the amplitudes of the two modes, respectively. 

Eq.~\eqref{eq:AmpFormMat} is applicable to the horizontal plane, too. However, the horizontal closed-orbit 
should include additional contribution from the beam energy variation due to the corrector kicks~\cite{BengtssondEE},
\begin{align}\label{eq:EnergyErr}
    \delta = \frac{\theta_1 D_1+\theta_2 D_2}{\alpha_c C}, 
\end{align}
where $\alpha_c$ is the momentum compaction factor, $C$ is the circumference, $D_{1,2}$ are dispersion function 
at the corrector locations. The corresponding terms in $D_P\delta$, where $D_P$ is the dispersion function at BPM $P$,
should be added to the amplitudes of the two horizontal modes.

The horizontal and vertical mode amplitudes  contain all 
linear optics information available from orbit measurements at the location. 
Instead of fitting the raw orbits to the lattice model, we can fit the amplitudes at all BPMs. 
Only 4 mode amplitudes per BPM are used to fit for 
linear optics, greatly reducing the size of fitting data. 
For the same large ring discussed earlier in this section, the Jacobian matrix has now only $3.4\times10^6$ elements, regardless 
of the number orbits used in the calculation. 
The same decomposition can be applied to cross-plane orbit data, which account for the linear coupling 
and the effects due to rotated correctors and BPMs. This will result in 4 additional mode amplitudes for each BPM. 

By choosing a proper excitation frequency,  i.e., to make $N\nu=$ integer, where $N$ is the number of 
sample points, 
the BPM data sample contain an integer number of periods, 
from which the amplitude of the cosine and sine  modes can be accurately computed for each BPM with
\begin{align}\label{eq:AmpFromData}
    A_c = \sum_{n=1}^N y(n) \cos 2\pi\nu n, \quad
    A_s = \sum_{n=1}^N y(n) \sin 2\pi\nu n.
\end{align}
If the excitation waveforms on the correctors are synchronized with data acquisition of the BPMs, 
the mode amplitudes can be readily calculated from the BPM data. 
If the correctors are not driven by fast waveforms, they can be scanned step by step, i.e., in the DC mode. 
This can be done at any storage ring  with two orbit correctors. 
In the AC mode, many orbits can be acquired in a short period of time. 
In the DC mode, fewer orbits may be read because data taking is slower, but with higher precision. 

\begin{figure}[thb]
\includegraphics[width=0.45\textwidth]{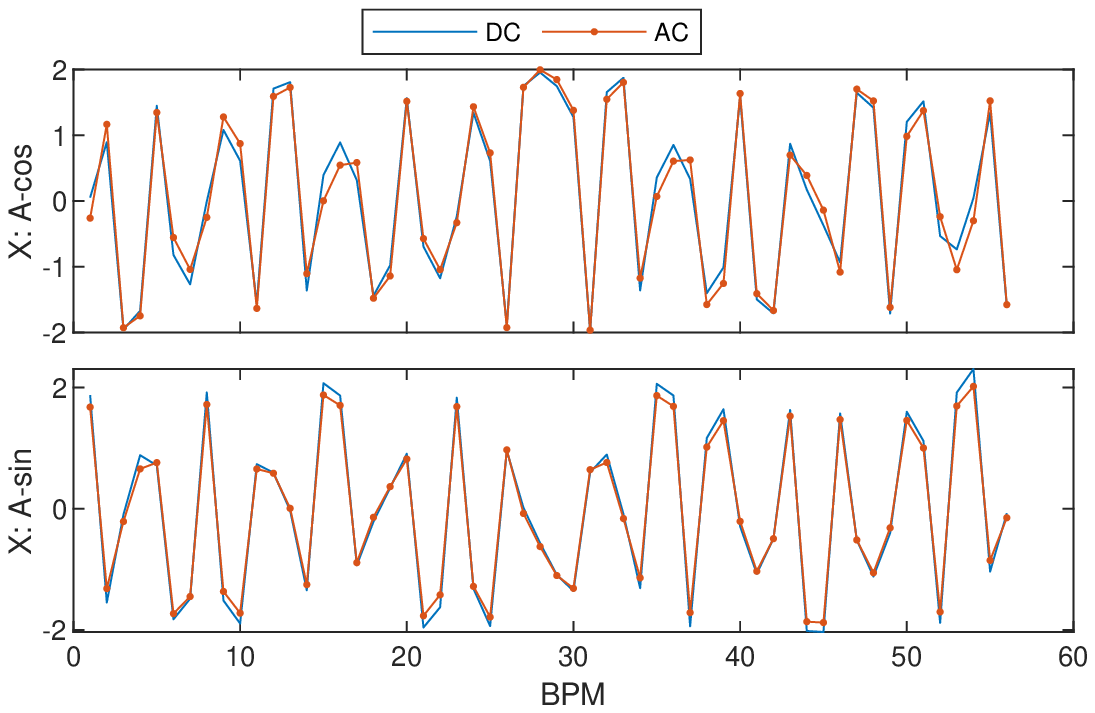}
\includegraphics[width=0.45\textwidth]{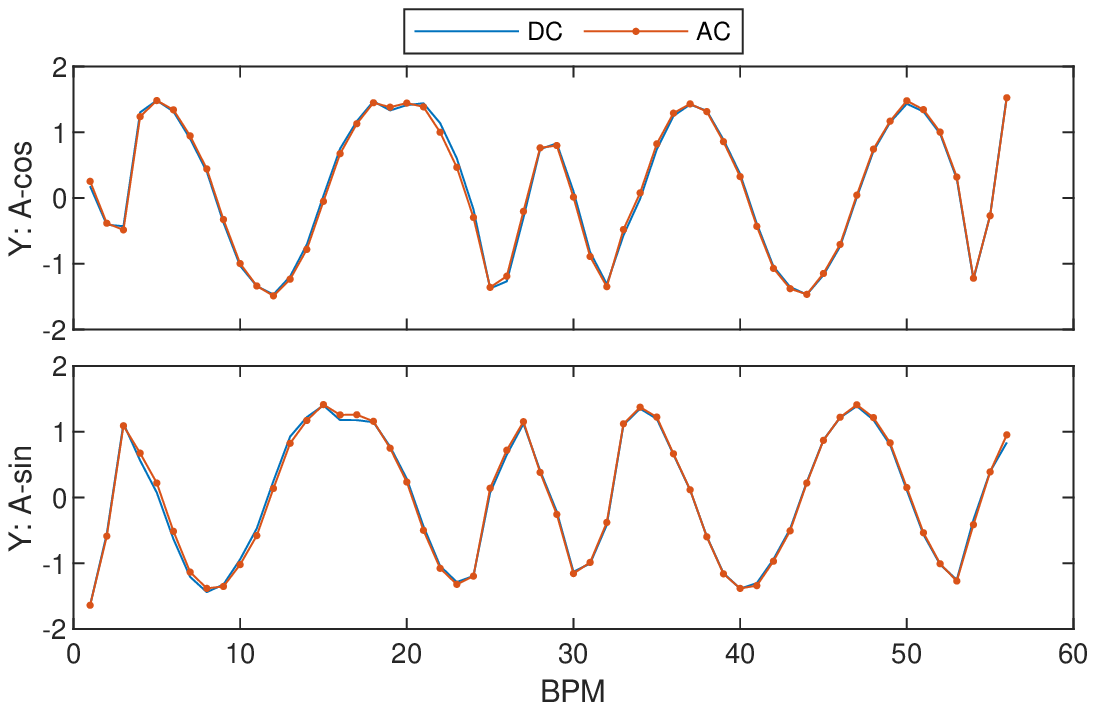}
\caption{Comparison of mode amplitudes for the horizontal (top) and vertical (bottom) 
planes for AC and DC LOCOM measurements on SPEAR3. The AC mode has 4 periods over 4000 data points taken over 1 second;
the DC mode scans one period with 180 data points. 
} 
\label{figACDCmodeAmp}
\end{figure}

Figure~\ref{figACDCmodeAmp} compares the amplitudes of the cosine and sine  modes for 
both the horizontal and vertical planes on the SPEAR3 storage ring. 
The AC LOCOM data has 4 periods in 4000 data points (as shown in Figure~\ref{figACorbitEx}) taken over 
just 1 second (for one plane). 
The DC LOCOM data contains 180 data points uniformly spaced in one period taken over about 2 minutes (for one plane). 
There is a good agreement between the AC and DC measurements. 

Using a lattice model, the mode amplitudes can be calculated with Eqs.~\eqref{eq:AmpFormMat} and \eqref{eq:EnergyErr}. 
An alternative is to calculate the closed orbit with the same corrector waveform in the lattice model 
and in turn to apply Eq.~\eqref{eq:AmpFromData} to the calculated orbits. 

By fitting the quadrupole strengths in the lattice model, the corrector gains, and the BPM gains, 
the differences between the measured and calculated amplitudes at all BPMs can be minimized in a 
least-square fashion. 
This is a standard practice in almost all methods for global linear optics correction. 
It is worth pointing out that degeneracy due to similar optics responses to changes in adjacent quadrupoles 
exists in this scheme, too, and the constrained fitting scheme can be used to alleviate the problem~\cite{HuangConstrainedLOCO}. 

\section{Application to NSLS-II \label{secAppNSLS2}}
Correction of linear optics with the improved LOCOM method has been successfully tested on the 
SPEAR3 storage ring and the NSLS-II storage ring. 
In the following we present results from the NSLS-II study as an example. 
The NSLS-II case is chosen because all its BPMs are capable of taking TBT data and it has kicker or pinger in both transverse planes, which offers a model 
independent verification of the effects of optics correction through the independent component analysis (ICA)~\cite{HuangICA05, YangICANSLS2}. 

At NSLS-II, the correctors can be set up to drive closed orbit motion with sinusoidal waveforms, 
as was demonstrated in AC-LOCO measurements~\cite{XiYangACLOCO}. 
However, the BPM data are not synchronized with the corrector waveforms, which complicates data processing 
for AC mode LOCOM measurements. 
For simplicity, we used DC mode LOCOM measurement in the optics correction tests. 
In the horizontal plane, the pair of driving correctors are %FHCM [2,1] and [3,1]
separated by a phase advance of $\psi_x=0.48\pi$ rad.
The phase advance between the vertical correctors is $\psi_y=0.35\pi$. 
The waveforms for the two correctors are in the form of~\cite{LOCOM}
\begin{align}
    \theta_1(n) &= \theta_{1m}\sin2\pi\nu n, \nonumber \\
    \theta_2(n) &= \theta_{2m}\cos(2\pi\nu n+\chi),
\end{align}
where the amplitudes $\theta_{1m}$ and $\theta_{2m}$ are related through 
$\theta_{1m}/\theta_{2m}=\sqrt{\beta_2/\beta_1}$, $\beta_{1,2}$ are beta functions 
at the corrector locations, and $\chi=\frac{\pi}{2}-\phi$. 
In the DC measurements, $N=120$ and $\nu=1/N$ are chosen, such that the orbit modulation 
complete exactly one period. 

\begin{figure}[thb]
\includegraphics[width=0.45\textwidth]{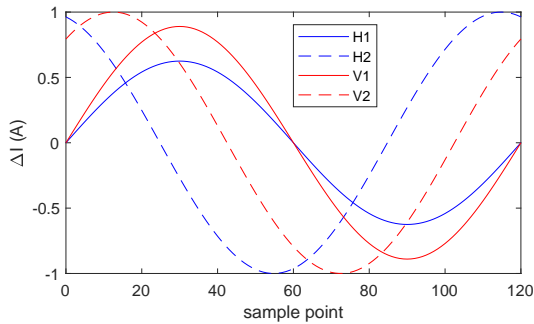}
\caption{
The corrector waveforms used in NSLS-II DC LOCOM measurements. 
Horizontal correctors are in blue; vertical correctors in red. Modulation amplitude is 1.0 A. 
} 
\label{figDCcorrWaveform}
\end{figure}

\begin{figure}[thb]
\includegraphics[width=0.45\textwidth]{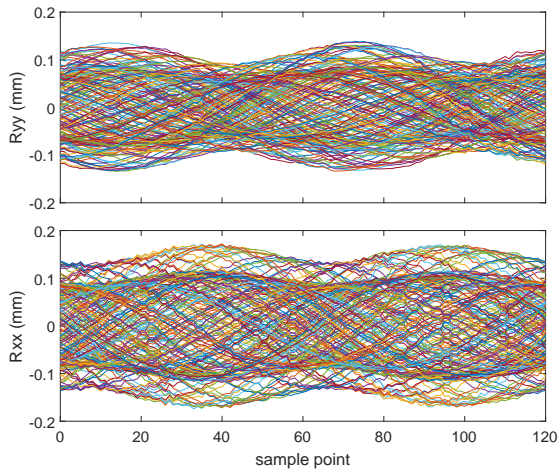}
\caption{
DC LOCOM data in an NSLS-II measurement. Top:  vertical;
bottom: horizontal. 
} 
\label{figRawDCLOCOM}
\end{figure}

\begin{figure}[thb]
\includegraphics[width=0.45\textwidth]{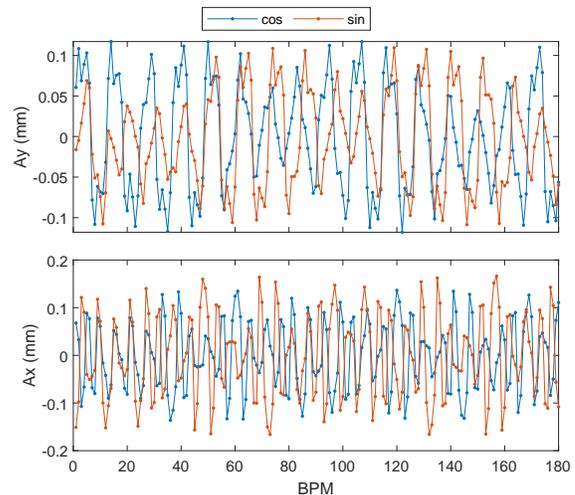}
\caption{
The horizontal (bottom) and vertical (top) mode amplitudes in the NSLS-II DC LOCOM measurement
as shown in Figure~\ref{figRawDCLOCOM}. 
%rmsY cos, sin:   0.0667    0.0588
%rmsX cos, sin:  0.0738    0.0913
} 
\label{figModeAmp}
\end{figure}

The corrector waveforms used in NSLS-II DC LOCOM measurements are shown in Figure~\ref{figDCcorrWaveform}.
The modulation amplitude in corrector current setpoint is up to 1 A.
An example of raw orbit modulation data for both planes on all 180 BPMs are shown in Figure~\ref{figRawDCLOCOM}. 
The mode amplitudes for this data set are plotted in Figure~\ref{figModeAmp}. 
In this case, the mode amplitudes contain the same linear optics information as the raw orbit data, but 
with only 1/60 of the numbers. 

NSLS-II has 30 double-bend achromat (DBA) cells. Six quadrupole magnets per cell is used 
as lattice fitting parameters, which are QL1, QL3, QH1, QH3, and two QM2 magnets. 
Including the BPM gains (360) and corrector gains (4), there are a total of 544 fitting 
parameters.

\subsection{Test case 1: changing  one  quadrupole}

\begin{figure}[thb]
\includegraphics[width=0.45\textwidth]{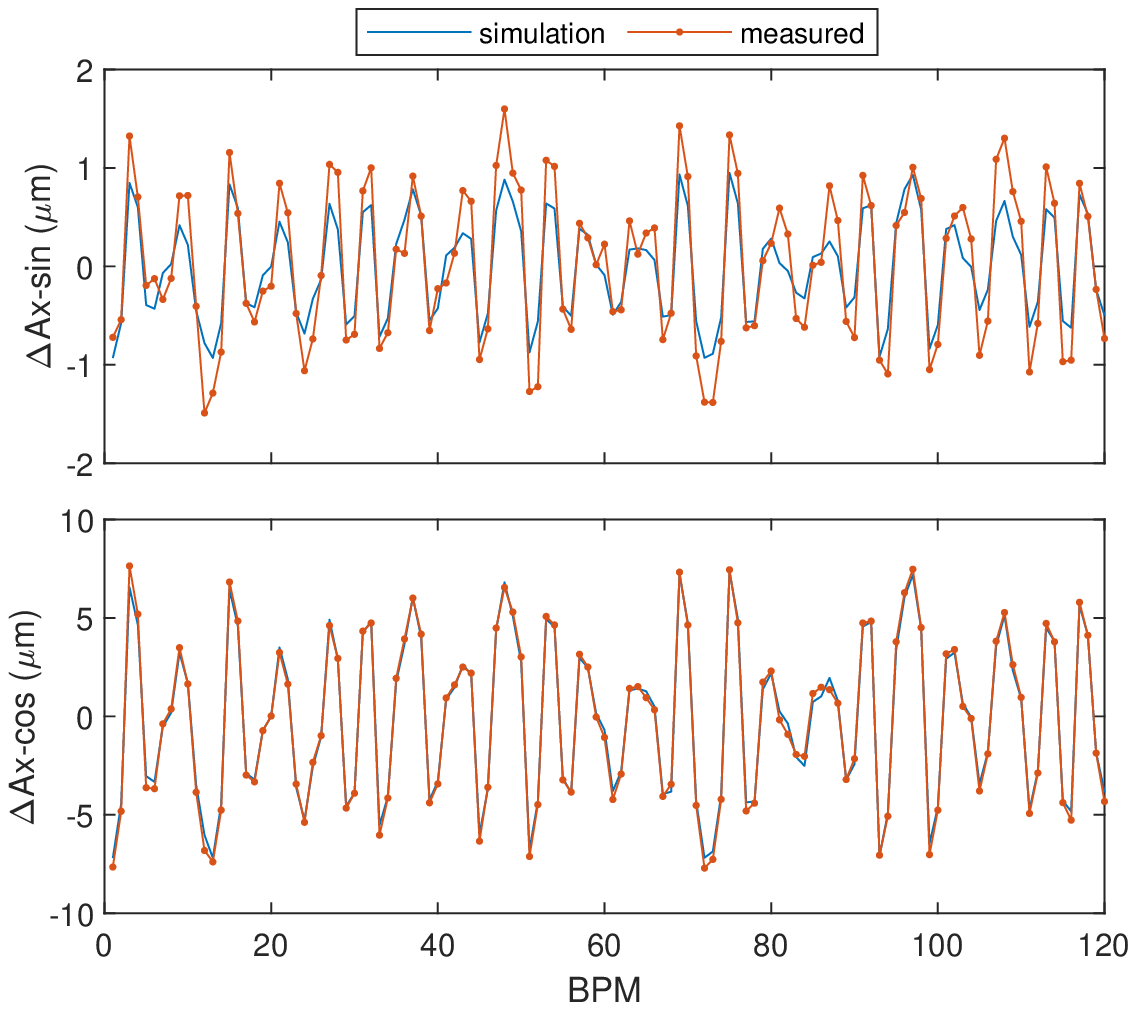}
\includegraphics[width=0.45\textwidth]{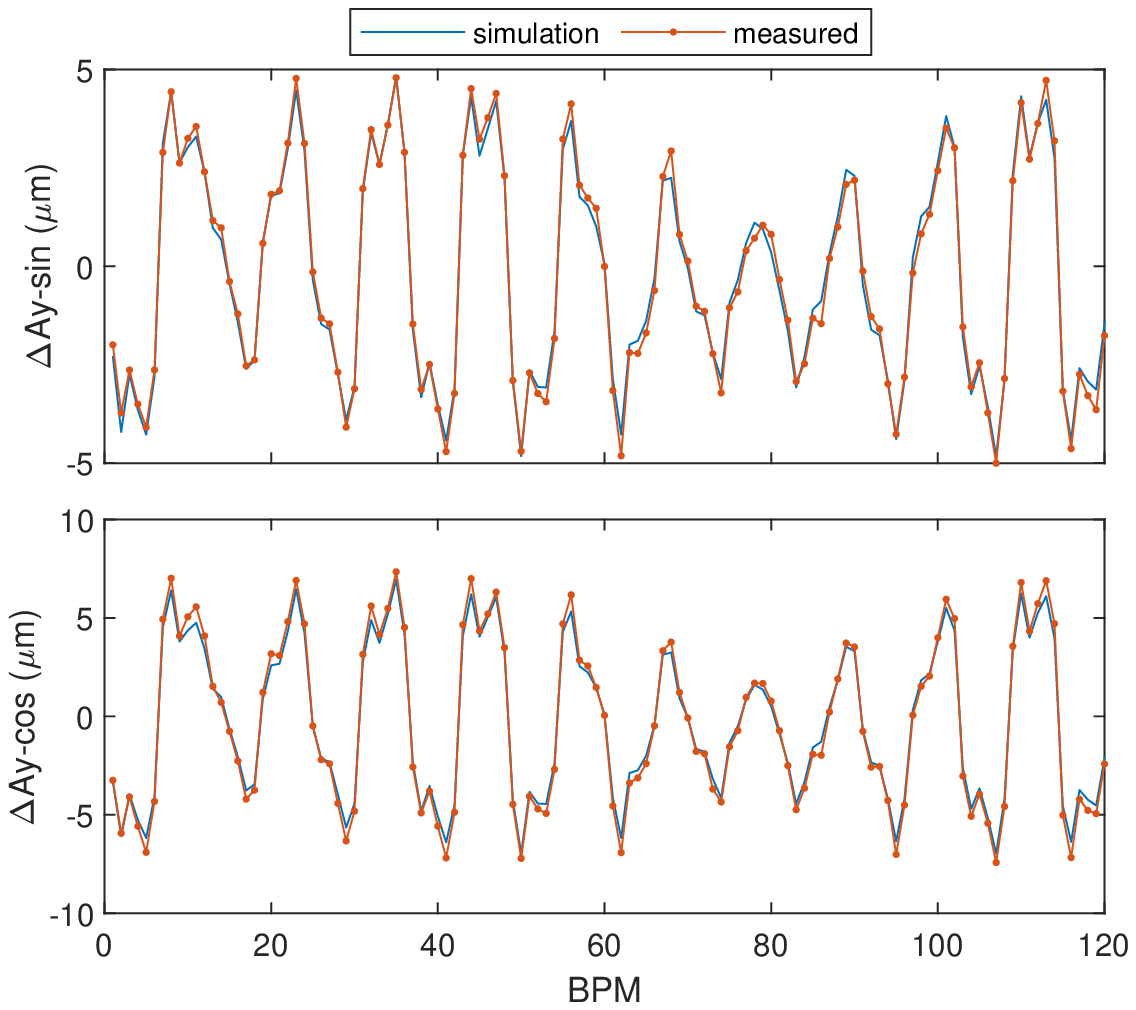}
\caption{
Changes to the cosine and sine mode amplitudes due to the change to the setpoint of one quadrupole 
magnet (QH3) by 1\%. 
Measured values and values from lattice model simulation are compared. 
Top to bottom: horizontal sine, horizontal cosine, vertical sine, and vertical cosine. 
} 
\label{figdiffCmpCSQH3n2}
\end{figure}

In one test case, the setpoint of one quadrupole magnet (the second QH3) in the NSLS-II ring 
was changed by 1\% to introduce linear optics errors to the machine. 
DC LOCOM data were taken before and after the changes were made to the machine. 
The changes to the quadrupoles cause distortions to the linear optics, which in turn change 
the LOCOM mode amplitudes for the same corrector waveforms. 
Figure~\ref{figdiffCmpCSQH3n2} shows the mode amplitude changes for both planes.
The measured changes are the differences of mode amplitudes from LOCOM measurements before 
and after the quadrupole was changed. 
The changes are also calculated with the lattice model by simulating the closed orbits 
and applying the same data analysis as done to the measured data. 
The simulated results are compared to the measurements. 
The good agreement between the measurement and the simulation indicates that the 
LOCOM data captured the features that represent the linear optics variations. 

\begin{figure}[thb]
\includegraphics[width=0.45\textwidth]{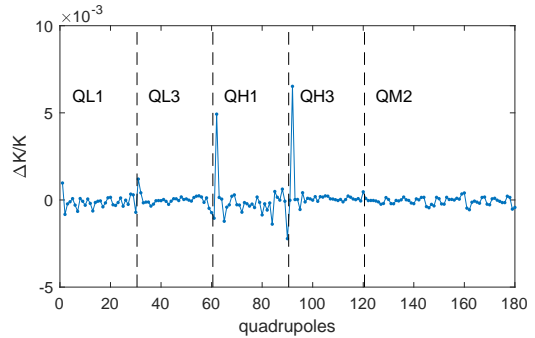}
\caption{
Differences of fitted $\frac{\Delta K}{K}$ between LOCOM data with a 1\% change 
to a QH3 magnet and the nominal condition. 
In the fitting results, nearby quadrupole magnets partially account for the induced error. 
} 
\label{figdKKDelta}
\end{figure}

The measured LOCOM data contain optics errors in the machine as well as the additional 
error due to the 1\% change to the QH3 magnet. Fitting the 
LOCOM data set to the lattice model will uncover both errors. 
However, the differences between the fitted quadrupole errors for the data set with the 
QH3 error and the nominal machine condition should reveal the effect of the single  magnet. 
Figure~\ref{figdKKDelta} shows the differences of the fitted $\frac{\Delta K}{K}$ from the LOCOM data for these two 
conditions. The fitting result found a $0.65\%$ error for the affected QH3 magnet 
and leakages to other quadrupole magnets in the vicinity, such as the second QH1 
magnet. Because of the degeneracy in the optics fitting problem~\cite{HuangConstrainedLOCO}, such leakages are not 
unexpected. 

\subsection{Test case 2: random changes to quadrupoles}
In a second NSLS-II experiment, lattice errors were introduced by making  small random changes to all quadrupoles.
The rms relative error sigma is 0.0073 for the 300 quadrupoles.
LOCOM data were taken and fitted to the model. 
The fitted quadrupole errors are then used to correct the linear optics errors on the 
machine. 
TBT BPM data were also taken, which provide an independent verification 
of the optics errors. 

\begin{figure}[thb]
 \includegraphics[width=0.45\textwidth]{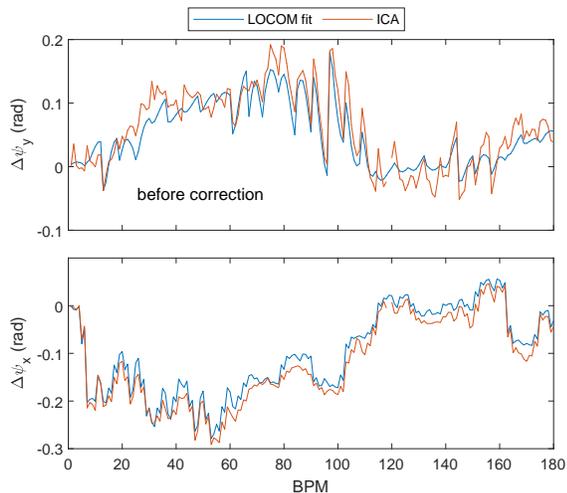}
\caption{Phase advance errors with initial random quadrupole errors 
obtained with TBT BPM data (``ICA'') or by fitting LOCOM data. 
} 
\label{figdiffCmpdPSIInit}
\end{figure}

\begin{figure}[thb]
 \includegraphics[width=0.45\textwidth]{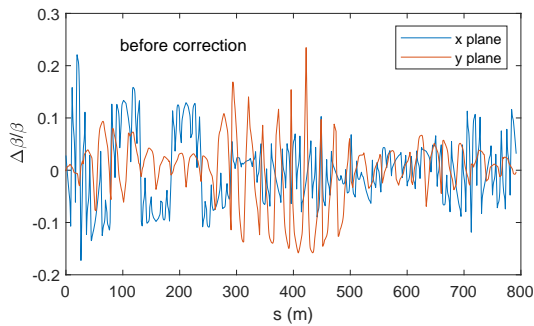}
\caption{Beta beating for the machine with initial quadrupole errors
obtained by fitting the LOCOM data. 
} 
\label{figBetaBeatIInit}
\end{figure}

\begin{figure}[thb]
 \includegraphics[width=0.45\textwidth]{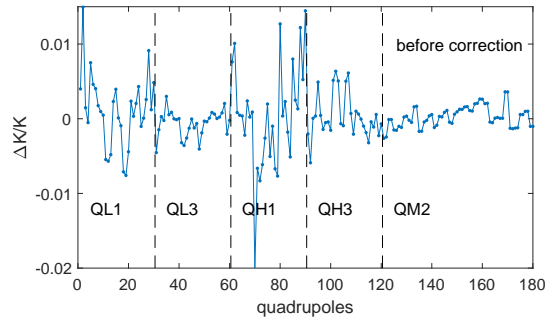}
\caption{The fitted $\frac{\Delta K}{K}$ by LOCOM for the machine condition with initial quadrupole 
errors. 
} 
\label{figdKKfittedIInit}
\end{figure}

Figure~\ref{figdiffCmpdPSIInit} compares the betatron phase advance errors 
obtained from TBT BPM data (processed with ICA) 
and by fitting the LOCOM data for the machine condition with the  
quadrupole errors. The phase advance errors here are the differences of the phase advances obtained by 
TBT BPM measurements 
or LOCOM fitting and that of the design lattice. 
The initial rms phase advance errors are 89.5~mrad (H) and 60.8~mrad (V) for the two planes, 
respectively, according to TBT BPM data processed with ICA.
There is a good agreement between the LOCOM and ICA for both planes. 
From the fitted lattice by LOCOM, we calculated the beta beating, which is shown in 
Figure~\ref{figBetaBeatIInit}. 
The rms beta beating is $6.7\%$ (H) and $6.1\%$ (V) for the two planes, respectively. 
The fitted quadrupole errors $\frac{\Delta K}{K}$ are plotted in 
Figure~\ref{figdKKfittedIInit}. 

\begin{figure}[thb]
 \includegraphics[width=0.45\textwidth]{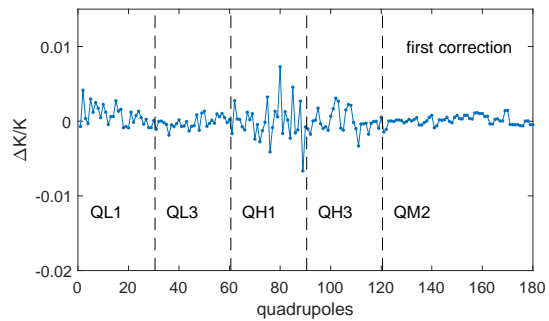}
\caption{The fitted $\frac{\Delta K}{K}$ by LOCOM for the machine condition after the first correction. 
} 
\label{figdKKfitted1stCorr}
\end{figure}

The fitted  $\frac{\Delta K}{K}$ were used to correct the linear optics errors 
on the machine. TBT BPM data and LOCOM data were then taken 
and processed in the same manner. 
The rms beta beating from the fitted lattice by LOCOM is 
$2.0\%$ (H) and $2.1\%$ (V) for the two planes, respectively. 
The fitted  $\frac{\Delta K}{K}$ (shown in Figure~\ref{figdKKfitted1stCorr}) from the first correction were applied to 
correct the optics for a second iteration.

\begin{figure}[thb]
\includegraphics[width=0.45\textwidth]{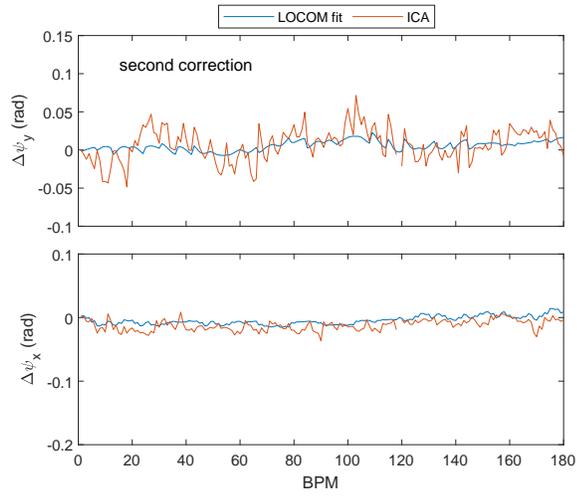}
\caption{Phase advance errors after the second correction 
obtained with TBT BPM data (``ICA'') or by fitting LOCOM data. 
} 
\label{figdiffCmpdPSIAfterCorr}
\end{figure}

After the second correction, the measured phase advance errors are 
very small, as shown in Figure~\ref{figdiffCmpdPSIAfterCorr}. 
The rms phase advance errors are 8.5~mrad (H) and 21.1~mrad (V) for the two planes, 
respectively, according to TBT BPM data processed through ICA. 
The rms beta beating becomes $0.78\%$ (H) and $0.95\%$ (V), respectively. 
The measured horizontal dispersion function is also brought closer to the design 
values, with rms errors at 13.7~mm before correction and 
7.6~mm after the second correction. 
Increasing the weight of the dispersion errors in the fitting setup 
may lead to better dispersion correction. 

\begin{figure}[thb]
 \includegraphics[width=0.45\textwidth]{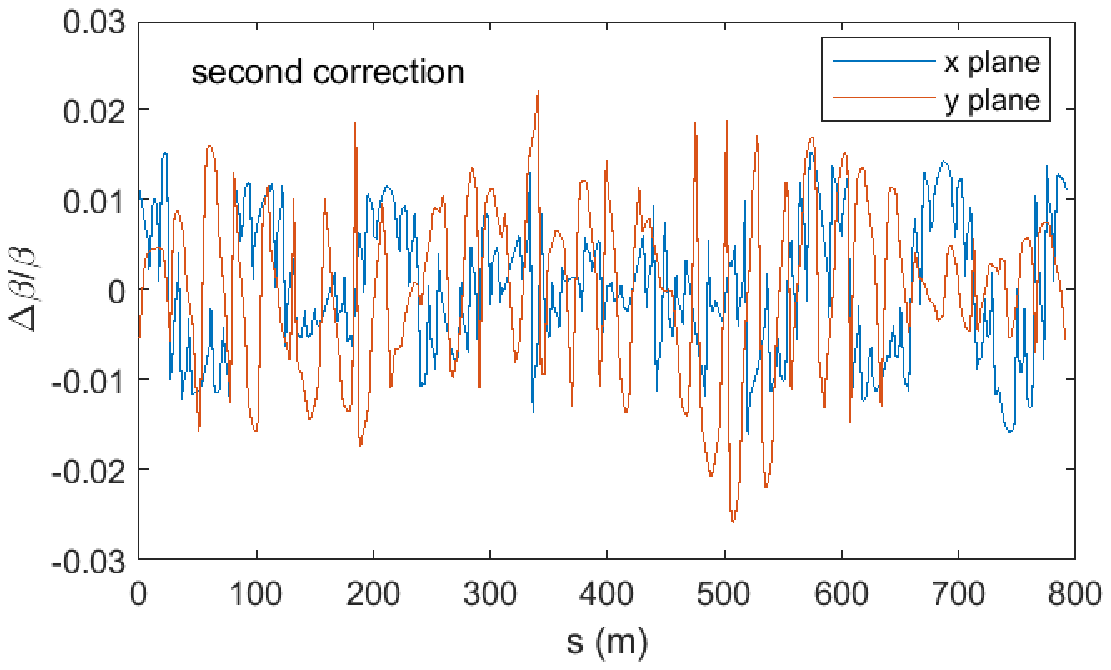}
\caption{Beta beating for the machine after the second LOCOM correction, 
obtained by fitting the LOCOM data. 
} 
\label{figBetaBeat2ndCorr}
\end{figure}

\section{Conclusion}
We have introduced an improved method to use closed orbit modulation data 
for linear optics correction and demonstrated the method experimentally on the 
National Synchrotron Light Source (NSLS)-II storage ring. 
Instead of fitting the individual orbits, as is done in LOCO~\cite{SAFRANEK199727} or the 
original LOCOM~\cite{LOCOM} method, 
the new method calculates the cosine and sine components of the orbit waveform at each 
BPM and fits only these four numbers (for the two transverse planes) to represent the linear optics information at the location. 
This approach greatly reduces the number of data points and allows a substantially 
larger collection of orbits to be used in fitting. 
This is especially useful for the case when the two modulating correctors are driven 
by fast waveforms and the BPM data are taken at the same frequency, where a large number of modulated orbits 
can be read in a short period of time (e.g., 10000 orbits in 1 second for NSLS-II). 
It is also a big advantage for large rings with many BPMs. 

The method can be easily extend to the case with cross-plane orbit data during 
corrector modulation, which can be used for linear coupling correction. 
In this case, 4 more mode amplitudes will be included as fitting data for each BPM. 
This will be tested in a future study. 

\begin{acknowledgments}
  This work was supported by the U.S. Department of Energy, Office of
  Science, Office of Basic Energy Sciences, under Contract No.
  DE-AC02-76SF00515 (SLAC) and Contract 
  No. DE-AC02–98CH10886 (BNL).  
\end{acknowledgments}

\clearpage
%\nocite{*}
%\bibliography{refs.bib}% Produces the bibliography via BibTeX.

\end{document}